\documentclass[twocolumn]{aastex631}

\defcitealias{Kluter2024}{Kl{\"u}ter \& Huston et al.}

\begin{document}

\title{Dazzle: Oversampled Image Reconstruction and Difference-Imaging Photometry for the Nancy Grace Roman Space Telescope}

\author[0000-0003-3316-4012]{Michael D. Albrow}
\affiliation{School of Physical and Chemical Sciences\\ University of Canterbury\\ Private Bag 4800\\ Christchurch\\ New Zealand}



\begin{abstract}

We present algorithms and software for constructing high-precision difference images to detect and measure transients, such as microlensing events, in crowded stellar fields using the Nancy Grace Roman Space Telescope. Our method generates difference images by subtracting an over-sampled reference, with iterative masking to address outlier pixels. We also provide an analytic correction for small dither offset errors. Microlensing event detection is achieved through a three-dimensional matched-filtering technique, optimized with Gaussian kernels to capture varying event durations, and verified through synthetic tests with high recovery rates. Transient photometry is performed via PSF fitting on difference images, using Nelder-Mead optimization for sub-pixel accuracy. The software, {\it Dazzle}, is available as an open-source Python package built on widely used libraries, offering accessible tools for the detection and characterization of transient phenomena in crowded fields.

\end{abstract}



\section{Introduction} \label{sec:intro}

The Nancy Grace Roman Space Telescope (formerly WFIRST, the Wide-Field Infrared Survey Telescope, and hereafter Roman) is expected to launch into L2 orbit in late 2026. 
The Galactic Bulge Time Domain Survey (GBTDS) is one of three of Roman's Core Community Surveys. Using Roman's Wide Field Instrument (WFI), it is expected to 
monitor approximately 2 deg$^2$ towards the Galactic Bulge at low galactic latitudes with a 15-minute cadence. The survey is expected to continue for 6 yearly seasons of $\sim 70$ days \citep{Penny2019, Wilson2023}.

Current plans are for a pipeline of direct PSF-fitting photometry using the methods of \citet{Anderson2000} from catalogued stars positions that will be established periodically through the duration of the survey.
We expect that the direct photometric pipeline will provide excellent-quality photometry for the majority of microlensing events with bright source stars. For faint and/or blended source stars, difference-image photometry is likely to provide 
better results.

 Difference imaging photometry \citep{Tomaney1996, Alard1998, Alard2000} has become the standard for Earth-based time-series imaging observations of crowded fields, and is used extensively for microlensing surveys of the Galactic Bulge \citep{Wozniak2000, Bond2001, Albrow2009}. A difference-imaging approach was also used for two notable HST time series campaigns, the WFPC2 observations of the core of 47 Tuc (GO-8267) in 1999 \citep{Gilliland2000, Albrow2001, Bruntt2001} and the ACS observations of Baade's Window in 2004 known as the SWEEPS project (GO-10475) \citep{Sahu2006}.

Terrestrial time-series imaging of crowded stellar fields usually have the following properties.
(i) The individual images, $T_k$, are spatially over-sampled, i.e. there are at least 2.5 pixels spanning the full-width at half maximum intensity of the point-spread function (PSF; 
the intensity function of the image of a star on the detector).
(ii) The width of the PSF is dominated by (variable) atmospheric seeing conditions.
Under these assumptions, we can define a difference image, 
\begin{equation}
    D_k = T_k - R \ast K_k - B_k,
\end{equation}
where $R$ is a reference image created from one or more individual images with the sharpest PSF, $\ast$ is the convolution operator, $K_k$ is a kernel that maps the PSF of $R$ to the PSF of $T_k$, and $B_k$ is the differential background.
The kernel, $K_k$, may be decomposed as a linear sum of analytic functions \citep{Alard1998} or as discrete pixel grid \citep{Bramich2013}. Depending on the representation, 
$K_k$ can account for subpixel offsets between $R$ and $T_k$.
The challenge in ground-based PSF difference-imaging photometry is to 
compute $K_k$ that minimizes the residuals in $D_k$.

In contrast, time-series images from space-based observatories like Roman have
a PSF that is constant in time, but the individual images are usually spatially
under-sampled, and have varying offsets (dithers) from each other. Difference-imaging photometry of these data requires completely different
methods than those employed for over-sampled images.
In this paper we present algorithms and software for a difference-image photometry pipeline for the Roman GBTDS. 

\section{Simulated data}

In the description of the methodology that follows, we use a set of simulated images from the forthcoming Galactic Bulge Time Domain Survey. 

To simulate the images, we first used the code 
{\it SYNTHPOP}\footnote{https://github.com/synthpop-galaxy/synthpop} (\citetalias{Kluter2024}~\citeyear{Kluter2024})
 to simulate
a population of stars in the direction of the Galactic Bulge centred on coordinates $(l, b) = (0, -3\deg)$ .
The specific model incorporated the Besançon galactic stellar population model \citep{Robin2003}, with the extinction map from \citet{Marshall2006} and the reddening law from \citet{ODonnell1994} (see also \citet{Cardelli1989}).

We then used this catalogue as input to the {\it RomanISIM}\footnote{https://github.com/spacetelescope/romanisim} code 
to simulate Roman Level-2 images. These are 4k $\times$ 4k images simulating the SCA-1 detector in the W146 filter. 
We used the {\it WebbPSF}\footnote{https://github.com/spacetelescope/webbpsf}
\citep{Perrin2014}
option within {\it RomanISIM} for the stellar PSF's. 
The simulated images have an integration time of 145.92 s and saturation of $\sim$ 2900 electron s$^{-1}$.
A total of 192 images were produced, at a cadence of 15 minutes, thus representing two days of data.

A short code, {\it RomanISim-simulate}\footnote{https://github.com/MichaelDAlbrow/RomanISim-simulate},
was used to control these steps. This code runs {\it SYNTHPOP} and converts the output to a format suitable for 
{\it RomanISIM}. It then runs {\it RomanISIM} to produce a set of images at the given observational cadence, introducing
the stellar proper motions from {\it SYNTHPOP} and random dithers. The script was configured to inject point-source point-mass-lens microlensing events with given parameters onto sets of random stars of given magnitude ranges.

\begin{figure}
    \includegraphics[width=\columnwidth]{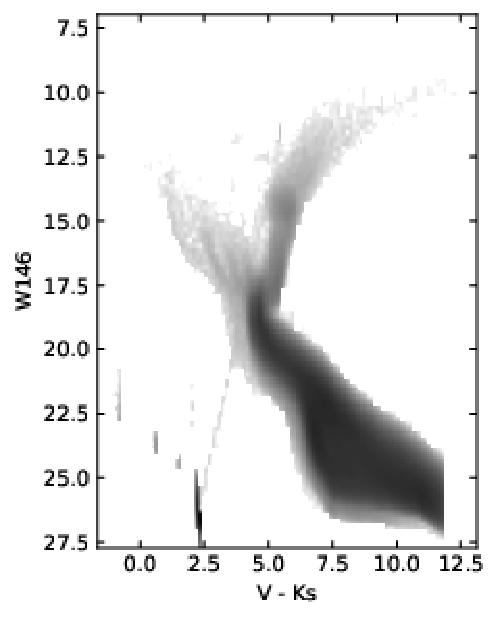}
    \caption{Hess diagram (shown in a log scale) of stars in the {\it SYNTHPOP} catalogue used as input to {\it RomanISIM}.}
    \label{fig:CMD}
\end{figure}

\begin{figure}
    \includegraphics[width=\columnwidth]{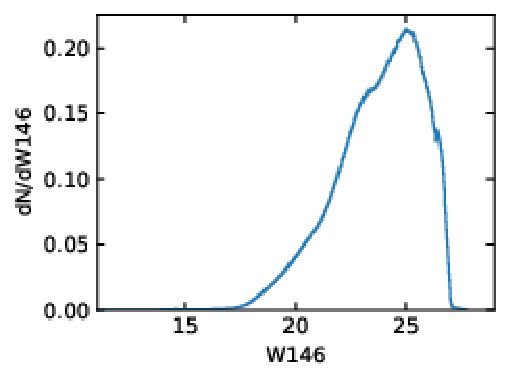}
    \caption{Luminosity function of stars in the {\it SYNTHPOP} catalogue used as input to {\it RomanISIM}.}
    \label{fig:LF}
\end{figure}

There are a total of 10.5 million stars in each of the simulated images. A Hess diagram and W146 (AB magnitudes) luminosity function of the 
of the input stars are shown in Fig.~\ref{fig:CMD} and Fig.~\ref{fig:LF} respectively.

\begin{figure*}[ht]
    \includegraphics[width=2\columnwidth]{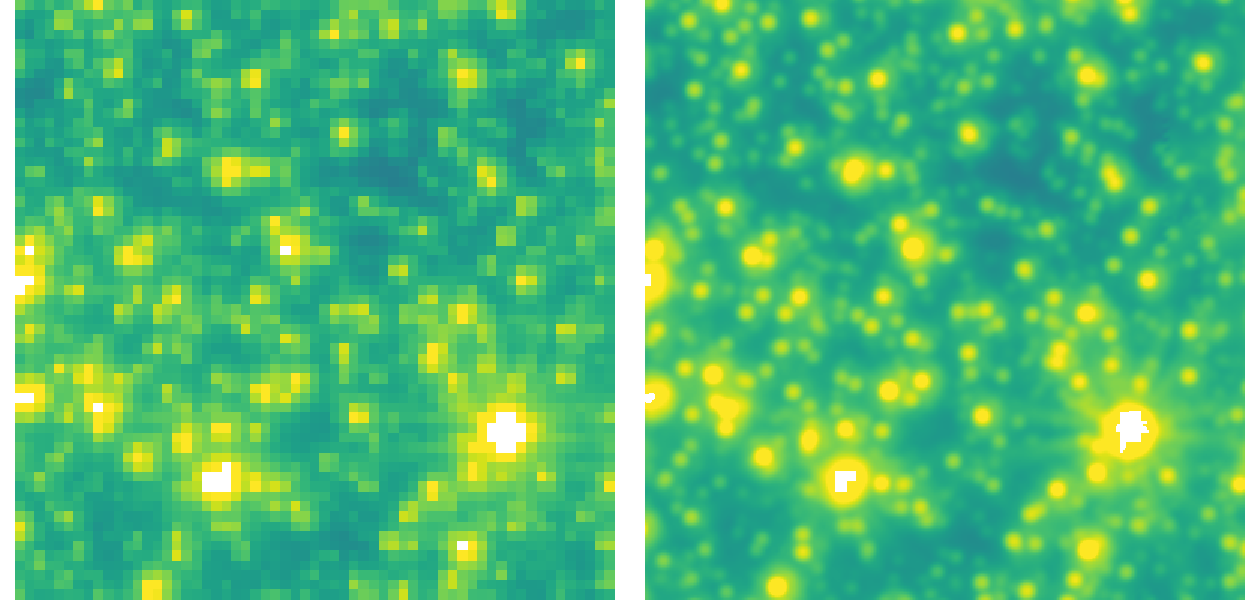}
    \caption{Sample 80 x 80 region of a simulated Roman image of the Galactic Bulge. Left panel: raw simulated image. Right panel: image at ten times the original sampling from a representation constructed from a stack of 192 randomly-dithered images. Panels have identical logarithmic intensity scaling and white pixels are saturated.}
    \label{fig:oversampled}
\end{figure*}

\section{Description of the Algorithm}

\subsection{Assumptions}

We assume that our raw data consists of a set of images, ${T_k(i,j)}$, of a single region of the sky. The images are spatially under-sampled and dithered. For now, we assume that the dithering consists only of linear translations between images, with no differential rotation. 
The number of images, and the dithering pattern are sufficient to sample the sub-pixel space in a sense that will be described below. There is a known over-sampled effective PSF that can be evaluated for any image location.

\subsection{Over-sampled image construction}
\label{sec:oversampled}

Our aim here is to construct, $R(x, y)$, an over-sampled representation of the observed scene, after convolution by both the telescope optics and the pixel response function. That is, the image of a single star should have the shape of the effect point-spread function (ePSF) as defined by \citet{Anderson2000}. 

As our reference grid, we adopt the pixel space of a single image, usually the first one, with indices defined at the pixel centres. The dithered offsets of each image from the reference are then determined, either from an astrometric solution found during previous processing (this will be the case for Roman) or some other means such as cross-correlation. Each pixel value in an image is a measurement of the over-sampled representation at its particular dither location. 

It is convenient in what follows to split each dither offset from the reference grid for image $k$ in the $x$ direction into an integer part, $\Delta x_k$, and a sub-pixel part, $\delta x_k$, with a similar definition for offsets in the $y$ direction.
Thus each image, 
\begin{equation}
    T_{k, ij} = R(x_i + \Delta x_k + \delta x_k, y_j + \Delta y_k + \delta y_k) + \epsilon,
\end{equation}
where $\epsilon$ means noise.

There are a number of potential representations that could be used to define the over-sampled image. For example
\begin{enumerate}
    \item
    A discrete array of pixels, defined at a higher spatial sampling than the original images. Interpolation between points on this grid could then be used to define the over-sampled image at any point.
    \item 
    A continuous two-dimensional analytic function, defined as a linear combination of basis functions that span the entire over-sampled image space. This would be a generalisation of the approach taken by \citet{Hogg2024} for one-dimensional spectra.
    \item
    A continuous spline representation, with knots defined on the boundaries of pixels in the reference grid.
    \item
    A series of two-dimensional analytic functions, again defined as linear combinations of basis functions that individually span the space of single pixels in the reference grid.
\end{enumerate}

\begin{figure*}[ht]
    \includegraphics[width=0.95\columnwidth]{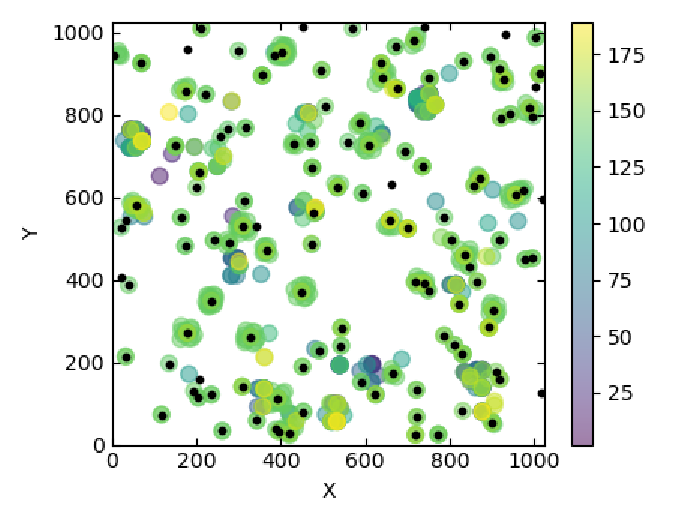}
    \includegraphics[width=0.83\columnwidth]{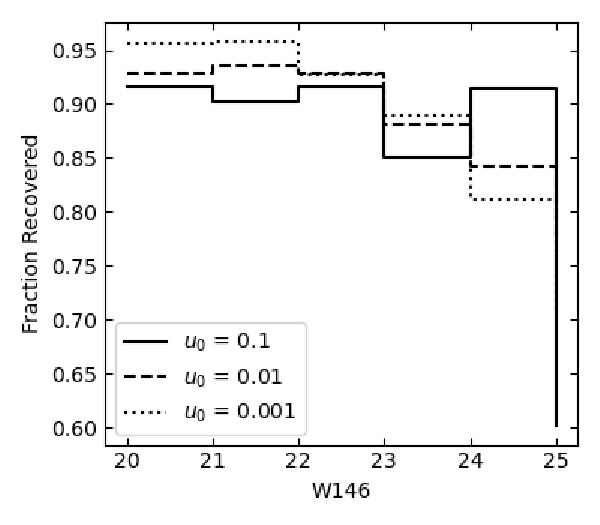}
    \caption{Left panel: Sample 1k $\times$ 1k region showing the locations of injected microlensing events (black dots) and 
    detected variables (coloured circles) using a gaussian kernel with $\sigma$ = 4.0 in the temporal direction and $\sigma$ = 1.0 in each of the spatial directions. The colour scale indicates the time of maximum in the detection-kernel-convolved image stack. The peak magnification of the injected events was at time 144 on this scale. Right panel: Overall recovery rate of injected microlensing events by magnitude and $u_0$. }
    \label{fig:detections}
\end{figure*}

\begin{figure*}[ht]
    \centering
    \gridline{
        \fig{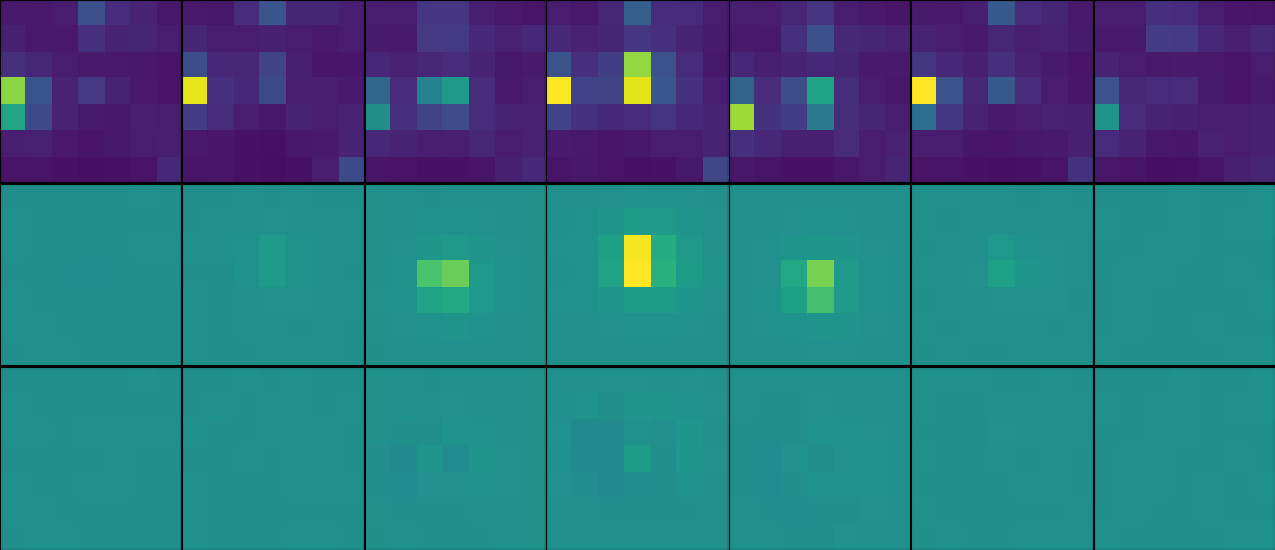}{0.48\textwidth}{(a) A relatively-isolated star, base magnitude W146 = 21.36. Top panel colour scale (0, 880) e$^-$ s$^{-1}$, lower panels colour scale (-64, 64) e$^-$ s$^{-1}$.}
        \fig{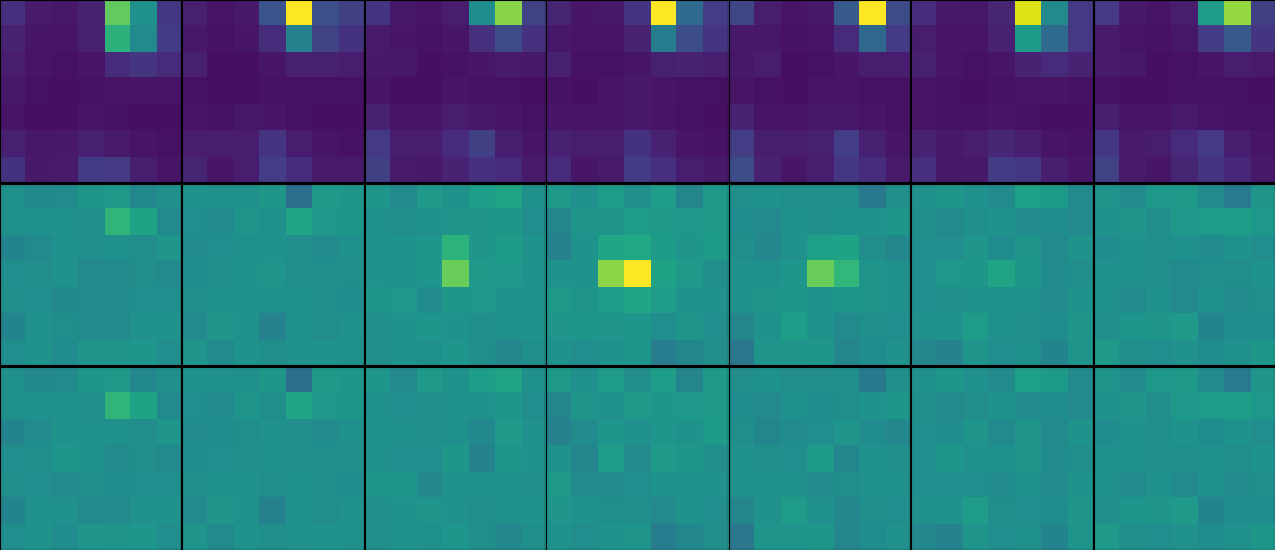}{0.48\textwidth}{(b) A faint isolated star, base magnitude W146 = 25.23. Top panel colour scale (0, 1130) e$^-$ s$^{-1}$, lower panels colour scale (-22, 22) e$^-$ s$^{-1}$.}
    }
    \gridline{
        \fig{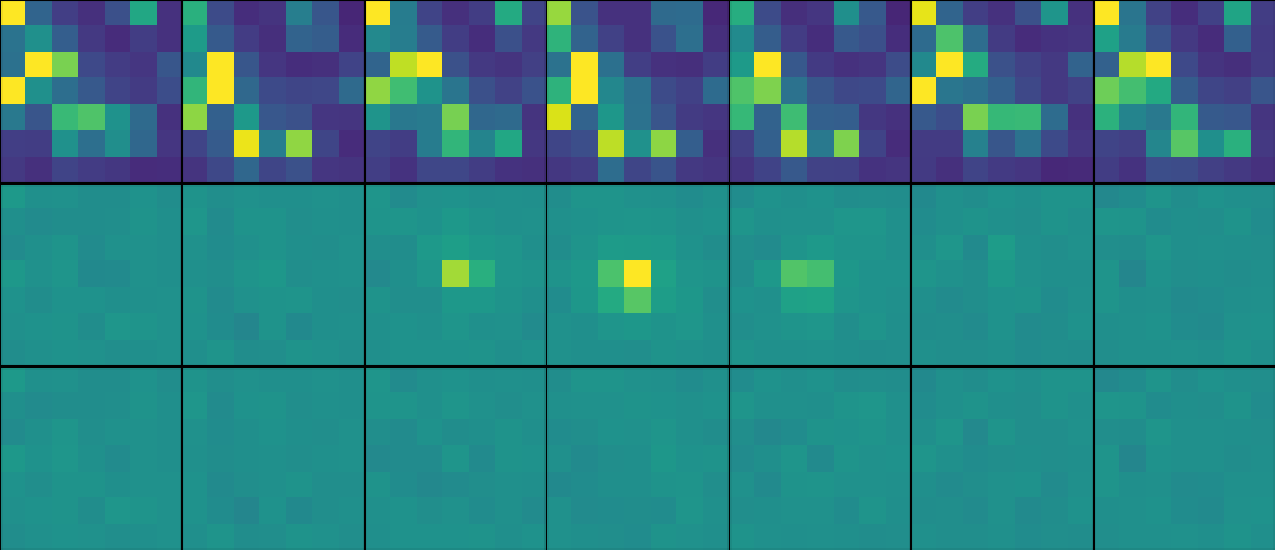}{0.48\textwidth}{(c) A faint star, base magnitude W146 = 24.24, in a crowded region. Top panel colour scale (0, 410) e$^-$ s$^{-1}$, lower panels colour scale (-50, 50) e$^-$ s$^{-1}$.}
        \fig{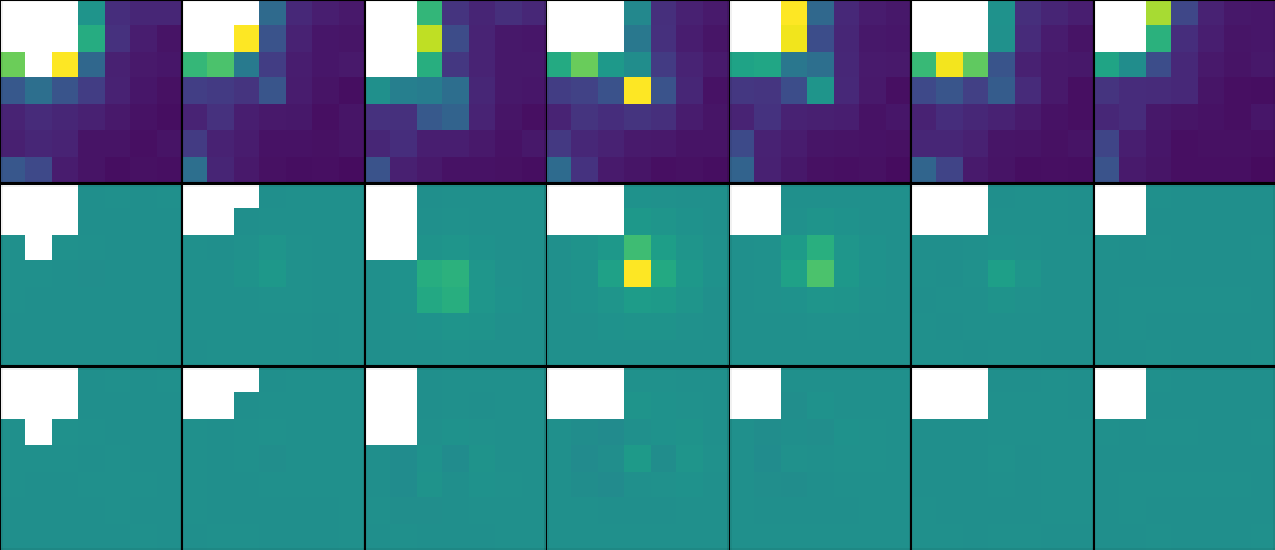}{0.48\textwidth}{(d) A brighter star, base magnitude W146 = 20.46, very close to a saturated star. Top panel colour scale (0, 2430) e$^-$ s$^{-1}$, lower panels colour scale (-2000, 2000) e$^-$ s$^{-1}$.}
    }
    \caption{Four examples showing time series of sample 7 x 7 pixel stamp images centered on a star. In each case
    a PSPL microlensing light curve with parameters $u_0 = 0.1, t_0 = 2160, t_E = 400$ minutes
    has been injected directly into the stellar flux before image simulation.
    The top row of each panel shows the direct simulated images. The central rows are difference images, and the bottom rows
    are the same as the central rows, but after subtraction of the fitted PSF. Pixels that are saturated in the
    direct images are coloured white. 
    Light curves from the complete sets of images are shown in Fig.~\ref{fig:light-curve}.
    The images shown here are at
    times 0, 1950, 2100, 2160, 2220, 2370, 2865 min.}
    \label{fig:time-series-images}
\end{figure*}

We prefer to have an analytic representation, rather than (i), in order to avoid interpolation at later processing stages. After some experimentation, we adopted the final option, (iv) above. Pragmatically, this is the solution that is computationally tractable (which is not the case for option (ii)), and mathematically and computationally more simple than option (iii). This was also the approach taken by \citet{Gilliland2000}. A further advantage is that the algorithm for over-sampled image generation described below is local for each pixel, and thus readily able to be parallelized. 

We define our over-sampled representation of pixel $(i,j)$ in the
reference grid as
\begin{equation}
R_{ij}(x,y) = \sum_{l=1}^{N} \sum_{m=1}^{N} \theta_{ijlm} B_l(x-x_i) B_m(y-y_j) , 
\end{equation}
where each $b_l(x)$ is a one-dimensional basis function of order $N$ spanning $-0.5 < x \leq 0.5$. We choose our basis set to be the Legendre polynomials, $\mathcal{L}_l(x)$. Since $\mathcal{L}_l(x)$ are orthogonal over $-1 < x \leq 1$, we use $B_l(x) \equiv \mathcal{L}_l(2x)$.

A small disadvantage of this basis is that it does not require the representation to be continuous across pixel boundaries in the reference 
grid. To mitigate this, the over-sampled representation can be made very smooth by extending the basis functions into small overlap regions with neighbouring pixels, in which case $B_l(x) \equiv \mathcal{L}_l(2x/f)$ for extension factor $f$.

From experimentation with simulated Roman images, we found
$N = 5$ with an extension factor $f = 1.2$, to work well for a smooth over-sampled representation and 
clean subsequent difference images.

We construct a design matrix, $X$, with elements
\begin{equation}
    X[k, l + Nm] =  B_l(\delta x_k) B_m(\delta y_k)
\end{equation}
for images $k$ with sub-pixel dithers $(\delta x_k, \delta y_k)$.
If $f > 1$, extra rows are appended to $X$ corresponding to the overlap pixel regions.
This matrix is common for all image pixels.

Then, for each pixel $(i, j)$ in the reference grid, we define a data vector, ${\mathbf{z_{ij}}}$, with elements
\begin{equation}
    z_{ij, k} = T_k(i + \Delta x_k, j + \Delta y_k) = T_{k,(i + \Delta x_k, j + \Delta y_k)}
\end{equation}
for images $k$, again supplemented by rows for neighboring pixels if $f > 1$. Note that we use the integer parts of the dithers in this expression.

Correspondingly, we define a data covariance matrix, $C_{ij}$, for each pixel, that is diagonal, with elements
\begin{equation}
    C_{ij, kk} = \sigma_{k,(i + \Delta x_k)(j + \Delta y_k)}^2,
\end{equation}
where $\sigma_{k,ij}$ is the uncertainty in $T_{k,ij}$.

For each pixel $(i, j)$ in the reference grid, the coefficients, $\theta_{ijlm}$, then have the standard linear algebraic solution,
\begin{equation}
    \theta_{ijlm} \equiv a_{ij(l+Nm)} = (X^T C_{ij}^{-1} X)^{-1} X^T C_{ij}^{-1} \mathbf{z_{ij}},
\end{equation}
that minimises
\begin{equation} \label{eqn:chi2}
    \chi^2 = \sum_{k} \frac{(T_{k,ij} - R_{kij})^2}{\sigma_{kij}^2},
\end{equation}
where $R_{kij}$ means $R$ evaluated at coordinates
$(x_i + \Delta x_k + \delta x_k, y_j + \Delta y_k + \delta y_k)$.

For $N$ = 5, there are 25 coefficients, $\theta_{ijlm}$, for each reference grid pixel $(i,j)$, so
formally, 25 is the minimum number of images that can be combined to construct $R$, and only if the dithering pattern well-samples the sub-pixel space. In practice, with random dithers, we have found
that 75 is a good minimum.

In Fig.~\ref{fig:oversampled} we show an example region of a single simulated Roman image, together with an over-sampled representation
constructed from 192 randomly dithered images.

\subsection{Difference images}

Once we have an over-sampled image representation, the process of creating difference images is straightforward. We define the difference image, $D_k$, corresponding to each original
image $T_{ij}$ as
\begin{equation}
    D_{ij} = T_{kij} - R_{kij}.
\end{equation}

However, our procedures described above assume a temporally-constant underlying scene, with all pixel value measurements distributed about their true values in a Gaussian sense with known variances. Real images 
contain pixels that violate these assumptions. Discrepant pixel values  can for example be due to cosmic rays, cosmetic detector defects, detector saturation, and variable stars. 

To mitigate these effects, we perform the above steps in an iterative fashion, using the difference images to detect and mask bad pixel values on an image-by-image basis, before constructing a new over-sampled representation. In practice, the algorithm converges satisfactorily in three to five iterations. We note that the algorithm that we have described for over-sampled image construction does not use any sort of interpolation of image data.

\begin{figure*}
    \centering
    \gridline{
        \fig{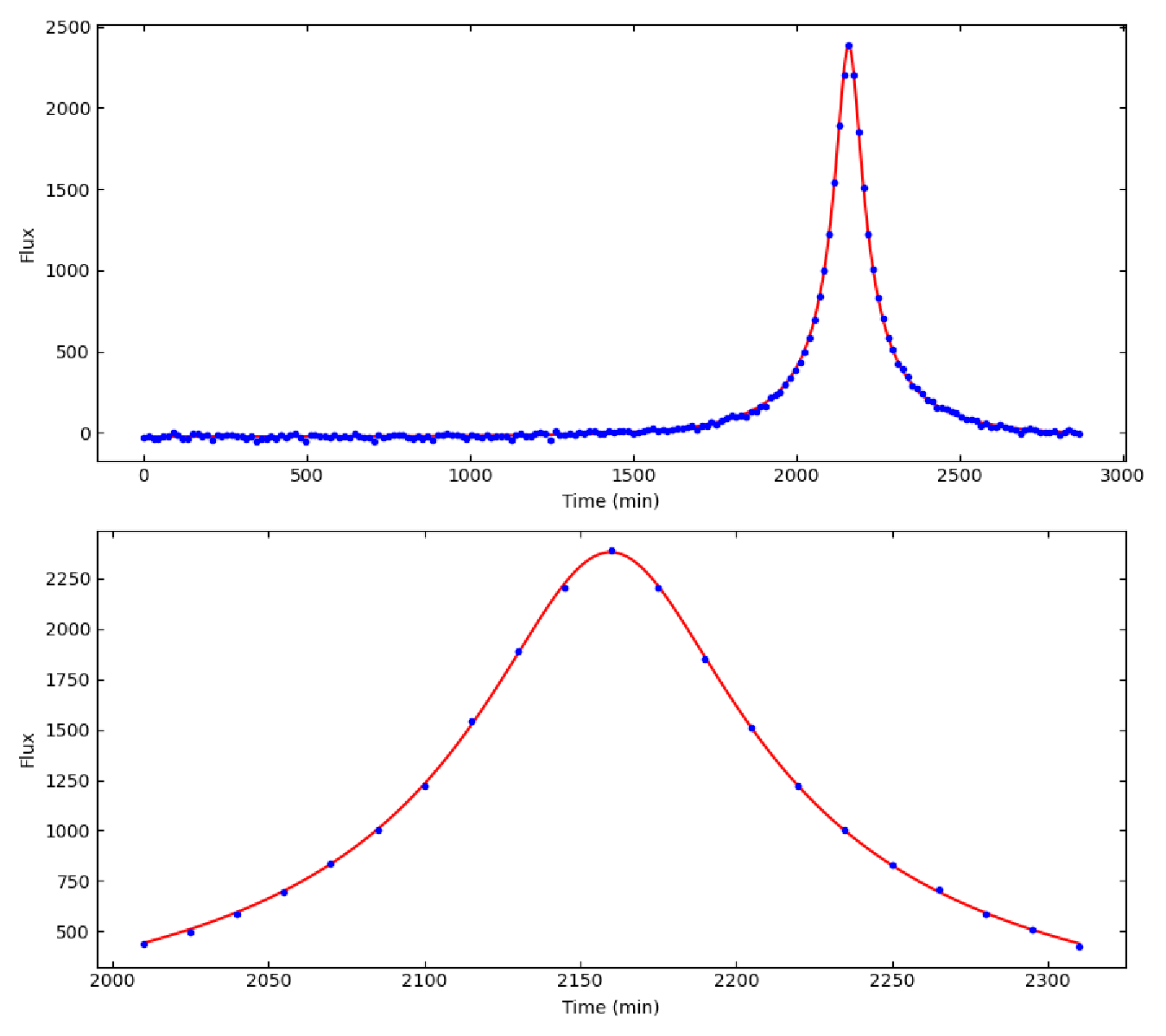}{0.45\textwidth}{(a)}
        \fig{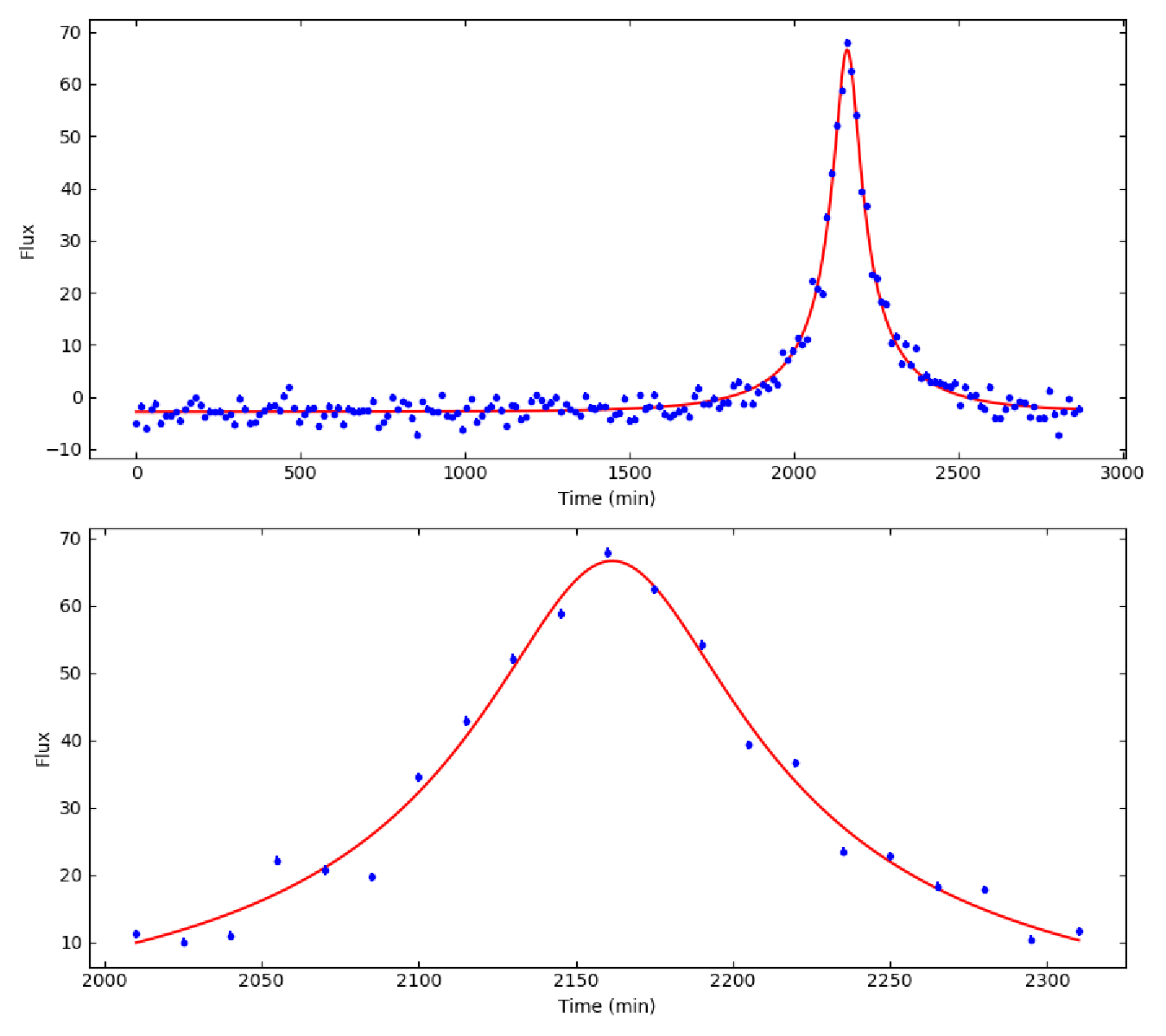}{0.45\textwidth}{(b)}
    }
    \gridline{
        \fig{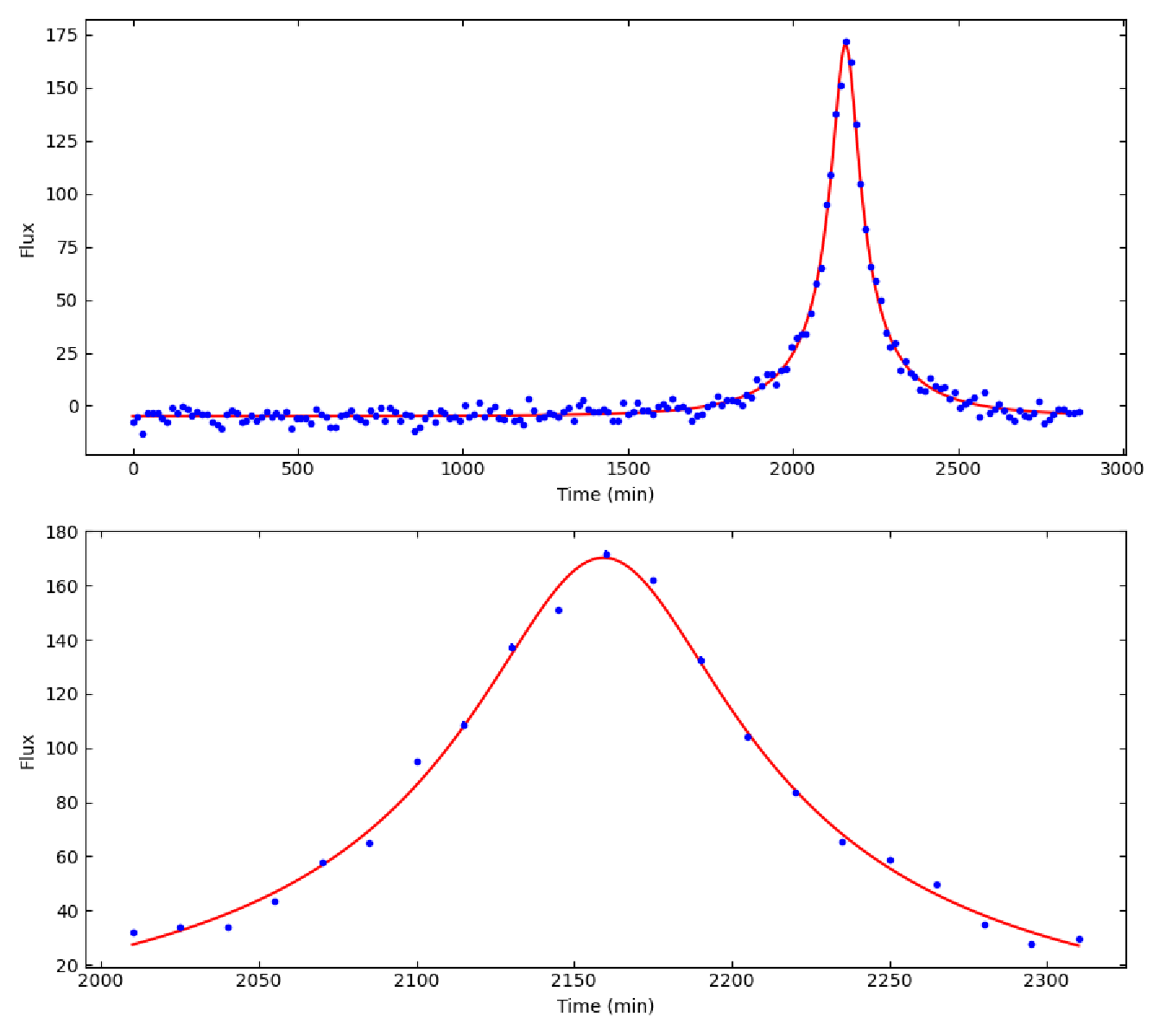}{0.45\textwidth}{(c)}
        \fig{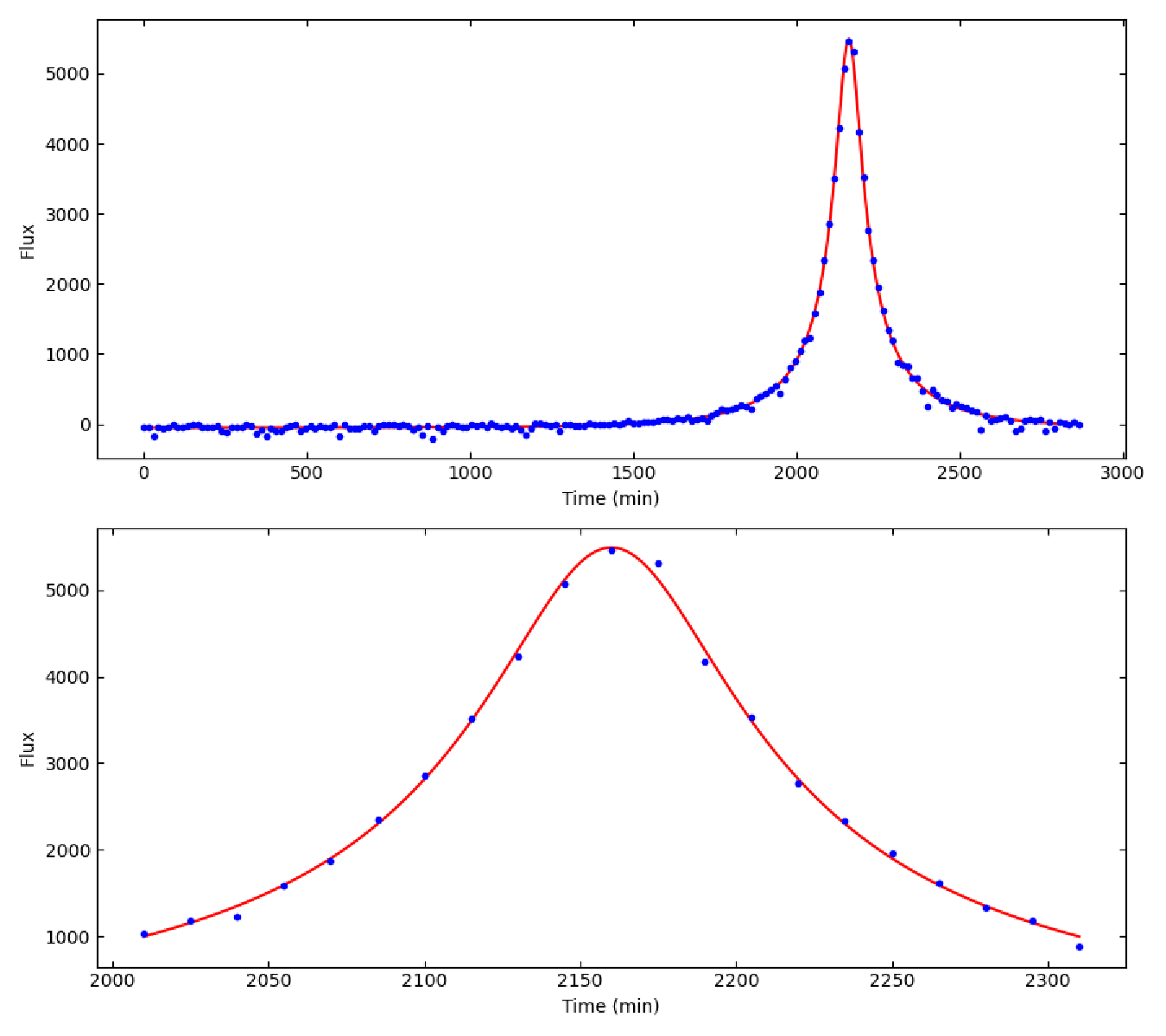}{0.45\textwidth}{(d)}
    }
    \caption{Complete light curves from PSF-fitting to the sets of simulated difference images shown in Fig.~\ref{fig:time-series-images}.
    The red lines are PSPL microlensing fits to the data.}
    \label{fig:light-curve}
\end{figure*}

\subsection{Correction of offsets.}

Assume that a single image, k, has a recorded dither position that is incorrect by a small amount. If $(\beta_k, \gamma_k)$ is the offset that has to be added to correct the dither position, then, assuming that the over-sampled representation is accurate, 
the corrected difference image will be
\begin{equation}
D'_{k, ij} =  R(x_i + \Delta x_k + \delta x_k + \beta_k, y_j + \Delta y_k + \delta y_k + \gamma_k) - T_{k,ij} + \epsilon.
\end{equation}
To first order, 
\begin{eqnarray} \label{eqn:R}
R(x_i + \Delta x_k + \delta x_k + \beta_k, y_j + \Delta y_k + \delta y_k + \gamma_k) = R_{kij} 
\nonumber \\
    + \beta_k \left. \frac{\partial R}{\partial x} \right|_{(x_i + \Delta x_k + \delta x_k, \, y_j + \Delta y_k + \delta y_k)} \nonumber \\
    + \gamma_k \left. \frac{\partial R}{\partial y} \right|_{(x_i + \Delta x_k + \delta x_k, \, y_j + \Delta y_k + \delta y_k)}.
\end{eqnarray}
so that
\begin{equation}
D'_{k, ij} = D_{k, ij} 
+ \beta_k   \frac{\partial R}{\partial x}
+ \gamma_k  \frac{\partial R}{\partial y} ,
\end{equation}
with the (analytic) derivatives evaluated at the locations from Equation~\ref{eqn:R}. Therefore, by minimizing
\begin{equation} 
    \chi^2 = \sum_{k} \frac{(D_{k, ij} + \beta_k \frac{\partial R}{\partial x}
    + \gamma_k \frac{\partial R}{\partial y})^2}
    {\sigma_{kij}^2},
\end{equation}
the offsets can be computed analytically as
\begin{equation}
    [\beta_k, \gamma_k]^T = \mathbf{A}^{-1} {\bf b},
\end{equation}
where

\begin{equation}
    \mathbf{A} = 
    \begin{array}{@{}l@{}}
    \left[
    \begin{array}{cc}
    \displaystyle\sum_{ij} \frac{1}{\sigma_{kij}^2} \left( \frac{\partial R}{\partial x} \right)^2 & 
    \displaystyle\sum_{ij} \frac{1}{\sigma_{kij}^2} \frac{\partial R}{\partial x} \frac{\partial R}{\partial y}  \\[10pt]
    \displaystyle\sum_{ij} \frac{1}{\sigma_{kij}^2} \frac{\partial R}{\partial x} \frac{\partial R}{\partial y} &
    \displaystyle\sum_{ij} \frac{1}{\sigma_{kij}^2} \left( \frac{\partial R}{\partial y} \right)^2
    \end{array}
    \right]
    \end{array},
\end{equation}

and
\begin{equation}
{\bf b} = - \left[
\sum_{ij} \frac{D_{k,ij}}{\sigma_{kij}^2} \frac{\partial R}{\partial x}, 
\sum_{ij} \frac{D_{k,ij}}{\sigma_{kij}^2} \frac{\partial R}{\partial y}
\right]^T.
\end{equation}

The algorithm can be applied iteratively, making small adjustments to the
individual image offsets at each iteration, before recomputing the oversampled representation. In practice this works well only for a small number of discrepant images with small dither errors, since it assumes that the initial over-sampled representation and its gradient are approximately correct.

\subsection{Microlensing event detection}

It is likely that most variable stars, including microlensing events, and stars with transiting exoplanets will be detected 
by various filtering methods from the primary direct-image photometry that will be made available from the pipeline data reduction
that is being developed by the Roman Galactic Exoplanet Survey Project Infrastructure Team (RGES-PIT). 

We are primarily concerned here with detection of microlensing events from the produced difference images. 
This might be the case for faint stars in crowded fields where the primary photometry is poor, or the source star is not present in the primary photometric catalogue.

Our approach is based on a matched filter. We begin by constructing a three-dimensional image stack, consisting of all of the difference images shifted by their integer pixel offsets of that they are approximately aligned. We then convolve the stack by a three-dimensional Gaussian kernel, which is symmetric in the two spatial directions, with a width similar to that of the stellar PSF. For the third (temporal) direction we use a number of trial widths in order to detect microlensing events of differing timescales.

We then search for peaks in the resulting three dimensional array using a local-maximum filter and requiring that detected peaks are above a configurable threshold. The threshold used depends on the temporal-direction kernel width. The detected peaks identify both the approximate spatial location and the time of peak brightness of candidate events.

As a test, we convolved the difference images with kernels with Gaussian $\sigma$'s of (2, 4, 8, 16) in the temporal direction. These numbers are in image units, with images at the 15-minute cadence. The injected microlensing events have a full-width at half-maximum flux of $\sim 3.5$ image units. In Fig.~\ref{fig:detections} we show an example
region of the detector space with injected events along with recoveries using a temporal $\sigma$ of 2. We define a recovery as being a detection within 3 pixels of the injected event. Most of the injected events are detected at approximately the correct epoch, along with a few false positives. 
The false positives are generally artifacts around saturated stars, due to the 
finite PSF size of 41 pixels used by WebbPSF.
Overall we find recovery rates of 90\% or better for sources with $W146 < 23$ and better than 80\% for 
$W146 < 25$. As expected, the recovery rate is generally higher for higher-peak-magnification events.
The recovery rate for these short-timescale events drops away as the temporal $\sigma$ increases.
The kernel parameters and detection thresholds could undoubtedly be further tuned to optimize the detection rate.

\subsection{Photometry}

For a given variable star (whether detected from the primary photometric light curves or from the difference images as discussed above),
we perform PSF-fitting photometry on the difference images to measure the difference-flux of the star at each epoch. 
We use a grid of ten-times-over-sampled PSF's that we generated with {\it WebbPSF} and, for each image, evaluate the PSF on a
square grid (stamp) with sides of length $2r+1$, with the observed image pixel scale. The evaluated PSF is centred on the assumed 
sub-pixel location of the star
in the difference image, accounting for the dithered offset of the image relative to the reference, and
the sub-pixel coordinates of the star in the reference.
The flux is then,
\begin{equation}
    F_k = \frac{
            \sum_{lm} P_{lm} D_{k(i+l)(j+m)} / \sigma_{k(i+l)(j+m)}^2
            }{
            \sum_{lm}  P_{lm}^2 / \sigma_{k(i+l)(j+m)}^2
            },
\end{equation}
with variance
\begin{equation}
    \sigma_{F_k}^2 = \frac{ \sum_{lm} P_{lm}^2 }{
            \sum_{lm}  P_{lm}^2 / \sigma_{k(i+l)(j+m)}^2
            },
\end{equation}
where $(i,j)$ is the integer pixel location of the centre of the star in difference image $k$, and 
$P$ is the grid-sampled PSF.

As shown by \citet{Albrow2009}, very accurate sub-pixel coordinates are need for precise PSF photometry. 
We use the Nelder-Mead method \citep{Nelder1965, Gao2012} to optimize the coordinates of the variable star by by minimizing
\begin{equation} 
    \chi_F^2 = \sum_{klm} \frac{(D_{k, lm} - F_k P_{lm})^2}
    {\sigma_{klm}^2},
\end{equation}
where $(l,m)$ range over the difference-image stamp
for a given $k$. Again $P_{lm}$ is evaluated from the PSF grid taking into account the stellar coordinates and integer and sub-pixel offsets of image $k$ from the reference.

In Fig.~\ref{fig:time-series-images} we show examples of the difference-imaging photometry process. The top row of images in each sub-figure are of a 7 x 7 pixel region
from selected images in the simulated set. Near the center of each of these images is a star (that 
may or may not be visible) that has had a PSPL microlensing event with $u_0 = 0.1$ injected into its flux at the time of image simulation. 
The microlensing event has a timescale of a few hours and would correspond to an isolated planetary-mass lens.
The left- and right-most images in each row show the field when the star is not magnified, and the five central images are the field at various epochs when the star is magnified. The central row of images in each sub-figure are difference images (with a smaller intensity scaling range) of the same parts of the field and same epochs as the top row. The magnified star is now clearly visible. The bottom row of the figure are the difference images (with the same scaling as the central row) after subtraction of the fitted PSF after optimization of the star's coordinates from the complete set of difference images.
Fig.~\ref{fig:light-curve} shows the full light curves for these events, measured from the complete sets of difference images. In all cases, the light curves show the expect smooth rise and fall, but we note that the (Poisson) error bars do underestimate the scatter. 

\begin{figure}[ht]
    \includegraphics[width=\columnwidth]{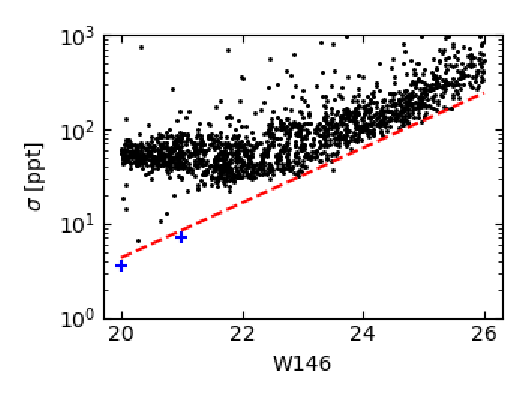}
    \caption{RMS photometric scatter, $\sigma$, in parts per thousand as a function of W146 magnitude. Each point corresponds to single light curve, with the scatter normalized to the baseline (unmagnified) flux in each case. The blue crosses are the total expected noise from the simulations of \citet{Wilson2023} (Fig 9 top panel, which are normalized to a 0.5 hour exposure time), and the red line is a linear fit to the lower envelope of points for which W146  $>$ 23. }
    \label{fig:scatter}
\end{figure}

In Fig.~\ref{fig:scatter} we show the RMS scatter of data points for each light curve about the fitted microlensing model. These are
normalized to the baseline (unmagnified) flux, so are an overestimate of the noise for magnified data points. We also compare these
measurements to those of the transiting exoplanet simulations of \citet{Wilson2023}. For brighter stars, there is a clear lower limit,
$\sigma \sim 0.03$ mag per data point in these difference-imaging photometry measurements. This is likely the regime where the 
direct PSF-fitting photometry will be superior. For fainter stars, the lower envelope of photometric precision can be 
represented by $\log_{10} \sigma [{\rm ppt}] = -5.19  + 0.29 \times$ W146.

\section{Implementation details}

The algorithms described above have been implemented in a python package that is available at \url{https://github.com/MichaelDAlbrow/Dazzle}. 
The package uses the standard {\it Numpy} \citep{Harris2020}, {\it Scipy} \citep{Virtanen2020} and {\it Astropy} \citep{Astropy2022} libraries.
Along with the core package, several python scripts are provided to reproduce the results from this paper.

The python code is deliberately written in a functional/procedural style, which seemed to better-suit the data flow for this project than a purely object-oriented approach.

\section{Summary}


We have presented algorithms and software for constructing high-precision difference images to facilitate the detection and photometric measurement of transients such as microlensing events in crowded stellar fields using the Nancy Grace Roman Space Telescope.

After constructing an over-sampled reference from dithered images, we generate difference images by subtracting the reference from each individual frame. To address discrepancies caused by outlier pixels, we apply an iterative procedure where pixels deviating significantly from the expected noise are masked and excluded in subsequent iterations. This enhances the quality of the resulting reference and difference images. For cases where dither offsets may be incorrect, we derive an analytic correction for small positional shifts, enabling a refined alignment.

For microlensing detection, we employ a three-dimensional matched-filtering technique, using a Gaussian kernel that is symmetric in the spatial plane and variable in the temporal direction to capture a range of potential event durations. Testing this method with injected synthetic microlensing events in the simulated dataset, we verified high recovery rates, particularly for bright sources with high peak magnifications.

Photometry of detected transient sources is performed on difference images using PSF-fitting methods to achieve high-precision flux measurements. By fitting a PSF model, generated with WebbPSF, and refining the sub-pixel coordinates using the Nelder-Mead optimization technique, we accurately determine the flux variations and deliver precise light curves for transient events. The difference-imaging PSF-photometry approach is particularly valuable in crowded fields where source blending can significantly degrade direct-PSF or aperture-based photometry.

Our implementation is encapsulated in the open-source Python package Dazzle, which is freely available on GitHub. Built on widely used libraries, including {\it Numpy}, {\it Scipy}, and {\it Astropy}, {\it Dazzle} provides accessible tools for the astronomical community to replicate and apply our methods. Its procedural design allows for straightforward adaptation and integration into diverse data reduction workflows, making it particularly well-suited for future surveys aiming to detect and characterize microlensing and other transient phenomena in crowded fields.


%



\software{
numpy \citep{Harris2020}, 
scipy \citep{Virtanen2020},
astropy \citep{Astropy2022} 
}



%

\bibliography{references}{}
\bibliographystyle{aasjournal}



\end{document}